\documentclass[aps,preprint,amsfonts,amsmath]{revtex4}

\usepackage{graphicx} %eps figures can be used instead
\begin{document}

\author{Martin J.~Greenall}
\affiliation{Institut Charles Sadron, 23, rue du Loess, 67034
  Strasbourg, France}
\affiliation{Department of Mathematics and Physics, Aberystwyth
  University, Physical Sciences Building, Aberystwyth, Ceredigion,
  SY23 3BZ, Wales, UK}
\author{Carlos M.~Marques}
\affiliation{Institut Charles Sadron, 23, rue du Loess, 67034
  Strasbourg, France}
\title{Can adding oil control domain formation in binary amphiphile bilayers?}
\begin{abstract}
Bilayers formed of two species of amphiphile of
  different chain lengths may segregate into thinner and thicker
  domains composed predominantly of the respective species. Using a
  coarse-grained mean-field model, we investigate how mixing
  oil with the amphiphiles affects the structure and thickness of the
  bilayer at and on either side of the boundary between two
  neighbouring domains. In particular, we find that oil molecules whose
  chain length is close to that of the shorter amphiphiles segregate
  to the thicker domain. This smooths the surface of the hydrophobic bilayer core on this side of
  the boundary, reducing its area and curvature and their
  associated free-energy penalties. The smoothing effect is weaker
  for oil molecules that are shorter or longer than this optimum
  value: short molecules spread evenly through the bilayer, while long
  molecules swell the thicker domain, increasing the surface area and
  curvature of
  the bilayer core in the interfacial region. Our results show that
  adding an appropriate oil could make the formation of domain boundaries
  more or less favourable, raising the possibility of controlling the
  domain size distribution.
\end{abstract}
\maketitle
\section{Introduction}
Bilayer membranes formed of a mixture of amphiphiles in solution can segregate
into domains of different compositions \cite {bagatolli}. Research in
this area has been driven by two major factors: the suggested role of
lipid domains in protein sorting in cell membranes \cite{brown_london} and the
capability of domain formation to control the surface properties of a
vesicle and to localise not only proteins
\cite{dumas,domanski,salamon} but also enzymes and particles \cite{bagatolli} within
its wall. Although membrane
bilayer domains have been most thoroughly investigated in lipid systems, recent work
has moved on to study mixtures of lipids and polymers
\cite{schulz,seitz,ruysschaert}, opening up the possibility of more
detailed control over bilayer properties such as stiffness, thickness
and hydrophobicity \cite{schulz}, and reinforcing the status of
membrane domains as an active and important field of research. 

The two different types of domain that form in a binary amphiphile
mixture may be in the liquid and gel phases respectively \cite{faller,tumaneng,hamada,zhao}, or may
both be in the liquid phase \cite{zhao,wu_mcconnell,veatch,garidel}, albeit with different degrees of internal
order in the amphiphile chains \cite{zhao,domanski}. Domain formation
can be controlled by a variety of factors, including the difference in chain
length between the two amphiphile species
\cite{domanski,faller,risbo,sanchez}, the lateral tension in the bilayer
\cite{hamada,akimov_pre} and the presence of a third species, such
as a protein or peptide
\cite{bagatolli,akimov_wetting,janosch,sperotto_mouritsen_enrich,jing},
cholesterol \cite{zhao,tumaneng,janosch,domanski},
ionised calcium \cite{knoll} or a ceramide \cite{westerlund}.

In this paper, we focus on a system in which two of these
factors interact, and use a coarse-grained mean-field
model to investigate how adding oil to a bilayer composed of two amphiphiles of different chain
lengths affects the structure of the membrane around the boundary
between two liquid domains. We have two main reasons for choosing this
problem, which also builds on our earlier work on oil droplets in bilayers
composed of a single amphiphile species \cite{gm_soft_matter}. First, from a
practical point of view, we wish
to find whether adding oil could prove
to be a viable technique for controlling domain formation and the
properties of the bilayer. Our second motivation is more theoretical. By choosing two amphiphiles that
differ only in length and an oil that is equally compatible with both, we
obtain a particularly simple system in which to study the addition of
a third species and its effect on bilayer conformation.

More specifically, we aim to find whether varying the size of the
added oil molecules can control the inhomogeneity that arises in the
membrane surface in the border region \cite{akimov_electro,kuzmin} and
its associated free-energy penalties. These determine how favourable
the formation of domain boundaries is, which in turn controls the size
distribution of domains \cite{akimov_electro}: if the free-energy cost
of forming a domain boundary is high, small domains will tend to fuse
together to form larger domains. We will quantify the inhomogeneity at
the domain boundary by calculating the changes in the surface area and
curvature of the hydrophobic bilayer core induced by changing the size
of the oil molecules.

The paper is organised as follows. In the next section, we
introduce the coarse-grained mean-field technique to be used, self-consistent field
theory (SCFT). We then present and discuss the results of our calculations,
and give our conclusions in the last section.

\section{Self-consistent field theory}\label{scft}

Self-consistent field theory (SCFT) \cite{edwards} has been used over
a number of years to model the equilibrium morphologies formed in
polymer melts and blends
\cite{maniadis,drolet_fredrickson,matsen_book}. It may also be extended to study
metastable structures \cite{duque,katsov1} and amphiphiles in solution
\cite{cavallo}, and has been applied to a wide range of
polymers, including homopolymers \cite{werner}, copolymers
\cite{mueller_gompper,wang} and mixtures of these \cite{denesyuk}. As
a mean-field model, SCFT requires less computer time than simulation
methods such as Monte Carlo, yet often yields predictions of the form
of individual structures that approach these more demanding methods in accuracy
\cite{cavallo,wijmans_linse,leermakers_scheutjens-shape}. Furthermore,
its simple, coarse-grained description of the polymer molecules will allow us to capture the
basic phenomenology of the system clearly.

We now give a short introduction to SCFT, and refer the interested reader to reviews
\cite{matsen_book,fredrickson_book,schmid_scf_rev} for an in-depth
presentation. A full description of our calculations for
amphiphiles in solution is presented in an
earlier paper \cite{gg}, and we give details only when the
current system differs from that described there. In SCFT, individual
molecules are modelled as random walks
in space, with the result that fine details of their packing and
structure are not taken into account \cite{schmid_scf_rev}. An ensemble
of many of these molecules is considered, and the inter-molecular
interactions are modelled by introducing contact potentials between
the molecules and assuming that the blend is incompressible \cite{matsen_book}. The
strength of the repulsion between the hydrophilic and hydrophobic species is specified by
the Flory parameter $\chi$ \cite{jones_book}. In order to reduce
the computational difficulty of the problem, a mean-field
approximation is then made
\cite{matsen_book}; that is, fluctuations are neglected. In the case of
long molecules, this
approximation is quantitatively accurate
\cite{cavallo,fredrickson_book,matsen_book}. Furthermore, SCFT can
provide useful qualitative insights when applied to
systems containing smaller
molecules, particularly lipid bilayers \cite{katsov1} and aqueous
solutions of copolymer \cite{schuetz}.

We now introduce the implementation of SCFT to our system of two amphiphiles and
oil in a solvent, which we model by a
mixture of two block copolymers with two incompatible homopolymers that
represent the oil and the solvent respectively. Although using a
mixture of polymers to represent a amphiphile-solvent system appears
slightly simplistic, models of this type have been used to capture the
broad phenomenology of a range of lipid and copolymer systems
\cite{katsov1,schuetz}. The mean-squared
end-to-end distance of the shorter copolymer is set to be $a^2N$, with $a$
being the monomer length and $N$ the degree of polymerisation
\cite{matsen_book}. One half of the
monomers in this polymer are hydrophilic (type A) and the other half are hydrophobic
(type B), so that the degrees of polymerization for the A and B blocks
are equal and $N_\text{A}=N_\text{B}$. For simplicity \cite{katsov1}, we also set the mean-squared
end-to-end distance of the A homopolymer solvent to $a^2N$. Since we
wish to focus on the effect of added oil on the structure of the
bilayer, we use a very long second copolymer, with $N_2\equiv\alpha N=16N$,
so that the inhomogeneity of the bilayer core becomes pronounced
around the domain boundary and can be easily studied.
We will consider
a wide range of oil sizes, and the degree of
polymerization $N_\text{O}\equiv \omega N$ of the oil will be varied between
$N/8$ and $4N$. Our focus on
bilayer structure and geometry also leads us to use oil molecules that
are composed of the same material as the hydrophobic B blocks, so that
the only interaction parameter $\chi$ in the system is that specifying
the strength of the repulsion between the A and B species.

In this paper, we fix the amounts of copolymer and homopolymer in the
simulation box; that is, we use the canonical ensemble. This
will make it easier for us to access more complex
structures such as segregated bilayers. Such structures are
more difficult to stabilise in ensembles where
the system is able to relax by
varying the concentrations of the various species, and can require
geometric constraints
to be applied to the density profile
\cite{katsov1}.

For completeness and to introduce the notation required for the
presentation of our results, we note
that the SCFT approximation to the free energy of our system is given by
\begin{align}
\lefteqn{\frac{FN}{k_\text{B}T\rho_0V}=\frac{F_\text{h}N}{k_\text{B}T\rho_0V}}\nonumber\\
& 
-(1/V)\int\mathrm{d}\mathbf{r}\,[\chi N
(\phi_\text{A}(\mathbf{r})+\phi_\text{A2}(\mathbf{r})+\phi_\text{S}(\mathbf{r})-\overline{\phi}_\text{A}-\overline{\phi}_\text{A2}-\overline{\phi}_\text{S})\nonumber\\
&
\times(\phi_\text{B}(\mathbf{r})+\phi_\text{B2}(\mathbf{r})+\phi_\text{O}(\mathbf{r})-\overline{\phi}_\text{B}-\overline{\phi}_\text{B2}-\overline{\phi}_\text{O})]
\nonumber\\
&
-(\overline{\phi}_\text{A}+\overline{\phi}_\text{B})\ln
(Q_\text{AB}/V)-((\overline{\phi}_\text{A2}+\overline{\phi}_\text{B2})/\alpha)\ln
(Q_\text{AB2}/V)\nonumber\\
&
-\overline{\phi}_\text{S}\ln(Q_\text{S}/V)-(\overline{\phi}_\text{O}/\omega)\ln (Q_\text{O}/V)
\label{FE}
\end{align}
where the $\overline{\phi}_i$ are the mean volume fractions of the
different components and the $\phi_i(\mathbf{r})$
are the local volume fractions. For the hydrophilic and hydrophobic blocks
of the shorter amphiphile, $i=\text{A}$ and
$i=\text{B}$ respectively, and for the corresponding blocks of the
longer amphiphile, $i=\text{A2}$ and $i=\text{B2}$. In the cases of
the oil and the solvent,
$i=\text{O}$ and
$i=\text{S}$ respectively. The Flory parameter, $\chi$, is set to
$15/N$, as using much larger values than this in conjunction with the
long species 2 copolymers could cause numerical instability. $V$ is the
total system volume, $1/\rho_0$ is the monomer volume, and $F_\text{h}$
is the SCFT free energy of a homogeneous system containing the same
components. The details of the individual polymers are contained in the
single-chain partition functions $Q_i$. These are computed
\cite{matsen_book} by
integrating over the
propagators $q$ and $q^\dagger$, which are also
used to calculate the polymer density
profiles \cite{matsen_book,fredrickson_book}. Due to the
fact that the molecules are modelled as random walks, the propagators
are calculated by solving modified diffusion equations with a field term that describes
the polymer interactions. These equations are solved in Cartesian
coordinates by a finite difference method
\cite{num_rec} with a step size of $0.04\,aN^{1/2}$. The dimensionless
curve parameter $s$ that specifies the distance along the polymer backbone is
taken to run from $0$ to $1$, and its step size in our finite
difference method is set to $1/1600$ for the long amphiphile species
and $1/400$ for the other species. We assume that
the system is translationally invariant along the $z$-axis, and so
consider an effectively two-dimensional problem in a rectangular
calculation box. The $x$-axis is taken to be perpendicular to the domain
boundary, and $x$ runs from $-L_x$ to $+L_x$, giving a box length of
$2L_x$. Similarly, the $y$-coordinate takes values from $-L_y$ to
$L_y$. In all calculations presented here, we set $L_x=14aN^{1/2}$ and
$L_y=4aN^{1/2}$, and impose reflecting boundary
conditions at all edges of the system.

The derivation of the mean-field free energy $F$ also generates a set
of simultaneous equations relating the fields $w_i(\mathbf{r})$ and densities
$\phi_i(\mathbf{r})$. To calculate the SCFT density profiles for a given set of
mean volume fractions $\overline{\phi}_i$, we make an initial guess
for the fields and solve the diffusion
equations to calculate the propagators and then the density profiles
corresponding to these fields. The new $\phi_i(\mathbf{r})$ are then
substituted into the simultaneous equations to calculate new values
for the $w_i$ \cite{matsen2004}, which are then used in turn to
calculate updated values for the $\phi_i$ by solving the diffusion
equation as described above. In order for the algorithm to remain
stable, the iteration must be damped, and we do not use these new
values of $w_i$ directly to calculate the $\phi_i$, but rather the linear combination
$\lambda w_i^\text{new}+(1-\lambda)w_i^\text{old}$ where
$\lambda\approx 0.04$. The procedure is repeated until
convergence is achieved.

The algorithm can be substantially accelerated by a simple
extrapolation procedure. This was developed by observing the typical
form of the error in the solution to the SCFT simultaneous equations \cite{matsen_book}
during the course of the iterations. To begin, we note that, with a
suitable initial guess for the fields (such as a broad potential well
in $w_\text{B}(\mathbf{r})$
for $x<0$ and a narrow one for $x>0$), the algorithm converges rapidly
to a set of density profiles with the general form of the segregated bilayer
we wish to study. However, the SCFT equations are not yet solved, and
display a sharp peak in their error terms at the boundary between the
two domains. The reason for this is that, although the density profiles have
the right overall form, the domain boundary has not yet been
correctly located. As the iterations are continued, the $x$-coordinate of
the boundary evolves towards its final value, and the error term peak
follows it, gradually decreasing in magnitude. In fact, the magnitude
of the error term peak proves to be approximately proportional to the
distance of the boundary from its final position along the $x$-axis. This allows us to
perform a simple linear extrapolation to estimate the final value of the
domain boundary. We then shift the fields $w_i$ along the $x$-axis by
a distance equal to the difference between the current and predicted
boundary positions. These shifted fields will then be used to continue the
iterations; however, we first need to deal with two technical issues. First, we note
that shifting the fields produces a region at one side of the system
where the $w_i$ are not known. Since the shift along the $x$-axis is
relatively small, we simply fill in the unknown region with the
values of the $w_i$ at the appropriate end of the unshifted
system, $w_i(\pm L_x,y)$. The shift will also have affected the
normalisation of the fields, which are usually defined
\cite{matsen_book} such that
$\int\mathrm{d}\mathbf{r}\,w_i(\mathbf{r})=0$. Appropriate constants
are calculated and added to the fields to correct this problem. This
extrapolation procedure need only be used once or twice during the
course of the iterations, and can reduce the error term rapidly. We
have also used this method to accelerate the convergence of SCFT
calculations on large vesicles \cite{gm_prl}, and it should generalise to a range of
density-functional problems involving an interface. 

\section{Results and discussion}\label{results}
In this section, we investigate the structure of the segregated bilayer for
a range of oil molecule sizes by studying the density profiles of the
various species. We then look in more detail at the surface of the
hydrophobic core of the membrane, and, in particular, at how its area
and curvature change as the size of the oil molecules is varied. Finally,
we study the effect on the shape and stability of the bilayer of
varying the oil concentration. To begin, we calculate the density profiles of segregated bilayers in a
system with volume fractions
$\overline{\phi}_\text{A}+\overline{\phi}_\text{B}=0.06942$,
$\overline{\phi}_\text{A2}+\overline{\phi}_\text{B2}=0.07246$ and
$\overline{\phi}_\text{O}=0.02036$. These values are chosen as they
allow the formation of two domains of approximately equal size. They
will be kept constant in the
first part of our study, although the length of the oil molecules will
be varied.

\begin{figure}[h]
\includegraphics[width=\linewidth]{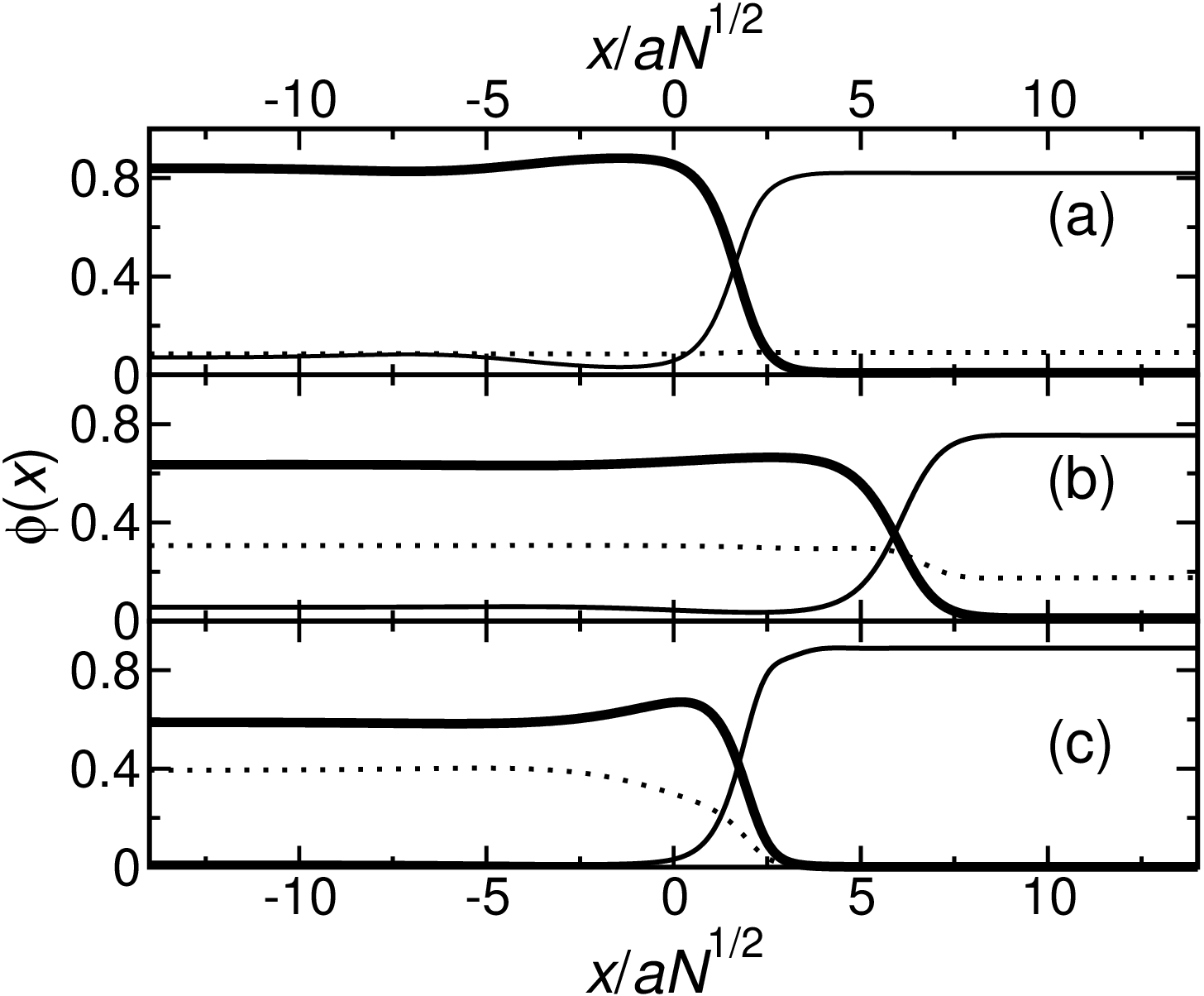}
\caption{\label{profiles_fig} Cuts through the density profiles in the
  bilayer core at $y=0$ for (a) $\omega=0.125$, (b)
  $\omega=1$ and (c) $\omega=4$. Thick full lines show the local volume
  fraction of the hydrophobic blocks of the larger amphiphile, thin
  full lines show the corresponding quantity for the shorter
  amphiphile, and dotted lines show the local volume fraction of the
  oil. The concentrations of the hydrophilic blocks are very low in the
  bilayer core and are omitted for clarity.} 
\end{figure}

In Figure
\ref{profiles_fig}, we plot cuts through the density profiles along the
$x$-axis at $y=0$. These run through the core of the membrane
perpendicular to the interface between the two domains. The first
point to note from Figure \ref{profiles_fig} is simply that solutions do
indeed
exist to SCFT with the form of segregated bilayers with two clear
domains separated by an interface. The domain containing mostly
longer amphiphiles is on the left of the interface, and that
containing mostly shorter amphiphiles is on the right. The
distribution of the oil molecules in the bilayer core depends strongly
on their size. In Figure \ref{profiles_fig}a, the oil
molecules are much shorter than either of the two amphiphile species,
with $\omega=0.125$. In consequence, they have no strong preference
for mixing with one amphiphile or the other, and spread evenly through
the two domains. In contrast, the oil molecules used in Figure
\ref{profiles_fig}b are longer, with $\omega=1$, and are the same size as
the shorter of the two amphiphiles. This means that they contain twice
the number of monomers as the hydrophobic sections of the shorter
amphiphiles, and mix less well with the right-hand side of the
bilayer. As a result, they are pushed over to the domain formed
predominantly of
longer molecules, which swells, moving the domain boundary to the
right. We note also that the concentration of oil molecules in both
regions is higher in Figure \ref{profiles_fig}b than in Figure
\ref{profiles_fig}a. This is because the longer oil molecules have a
stronger repulsive interaction with the solvent, as increasing
$\omega$ increases the product $\chi N$, which determines the
interaction strength \cite{jones_book}. In Figure \ref{profiles_fig}c, the
oil molecules are still longer, with $\omega=4$, and mix hardly at all
with the shorter amphiphiles. However, the swelling of the
domain perpendicular to the domain boundary seen in Figure \ref{profiles_fig}b
is absent, and the interface has returned to a position close to the
centre of the system, as in Figure \ref{profiles_fig}a. A natural
explanation for this is that the left-hand side of the bilayer has
swollen perpendicular to the plane of the membrane; that is, it has
become thicker. We will discuss this point in more detail later.

\begin{figure}[h]
\includegraphics[width=\linewidth]{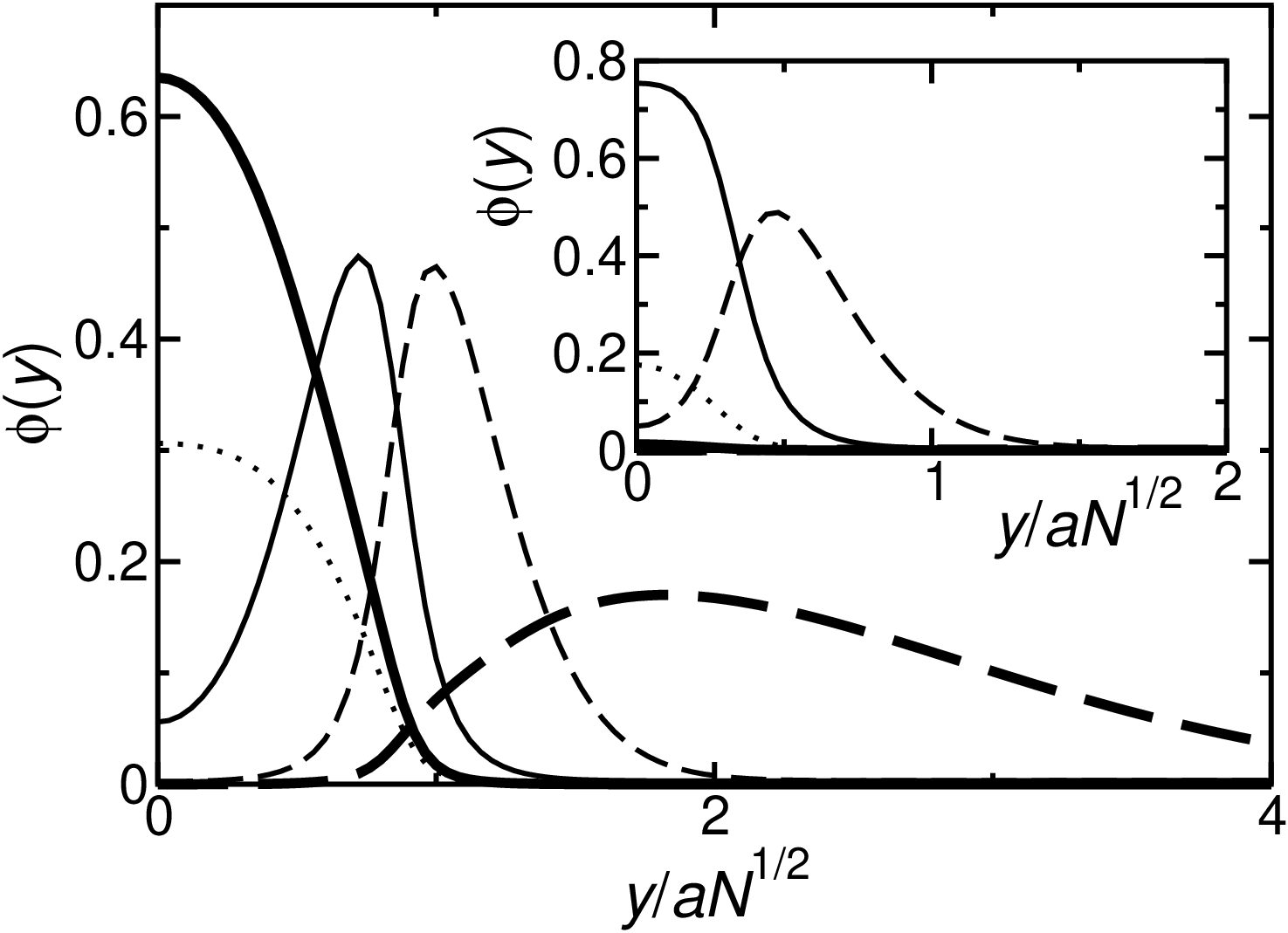}
\caption{\label{centre_edge_fig} Cuts through the density profiles at
  $x=-L_x$ (main panel) and $x=+L_x$ (inset). Thick and thin dashed lines show the
  local volume fractions of the hydrophilic components of the long and short amphiphiles
  respectively, and the other species are labelled as in Figure \ref{profiles_fig}.} 
\end{figure}

To give some more insight into the membrane structure, we plot cuts
through the density profiles of the various species in the bilayer at
the left- and right-hand sides of the system ($x=\pm L_x$) in the
direction ($y$)
perpendicular to the plane of the membrane. We focus on the case where
$\omega=1$, the system shown in Figure \ref{profiles_fig}b. In the main panel of
Figure \ref{centre_edge_fig}, we show the density profiles on the left of
the system ($x=-L_x$), where the bilayer is formed
predominantly of the longer amphiphile species. We see that the
structure of the bilayer is more complex than might at first have been
expected. Although the core of the membrane is indeed composed mainly of the
longer amphiphile species, there is a thinner layer of the shorter
amphiphile on the outside of the bilayer, at $y\approx aN^{1/2}$. At
the other side of the bilayer ($x=+L_x$), shown in the inset, the structure is simpler, and the
bilayer is formed almost exclusively of the shorter species. This
shows that the segregation of amphiphiles due only to a difference in size
between the two molecular species is far from perfect, with the
bilayer formed of the shorter amphiphiles splitting into two leaflets
at the domain boundary and coating the outer surface of the thicker domain.

We now proceed to study the effect of the oil molecular size on the
structure of the bilayer core in more detail. To this end, we plot the
interface between the hydrophobic core and its hydrophilic
surroundings, defined as the locus of the points where
$\phi_\text{B}(x,y)+\phi_\text{B2}(x,y)+\phi_\text{O}(x,y)=0.5$. For
clarity, and to help our later analysis of the bilayer shape, which will
involve the calculation of derivatives of the core outline, we fit our
discrete SCFT results with a curve of the form
\begin{equation}
y(x)=a_0+\frac{a_1}{1+\exp[-(x-a_2)/a_3]}+a_4\exp[-(x-a_5)^2/a_6]
\label{boltzmann_gaussian}
\end{equation}
where the $a_i$ are adjustable parameters. This formula gives an
excellent fit to the data.

\begin{figure}[h]
\includegraphics[width=\linewidth]{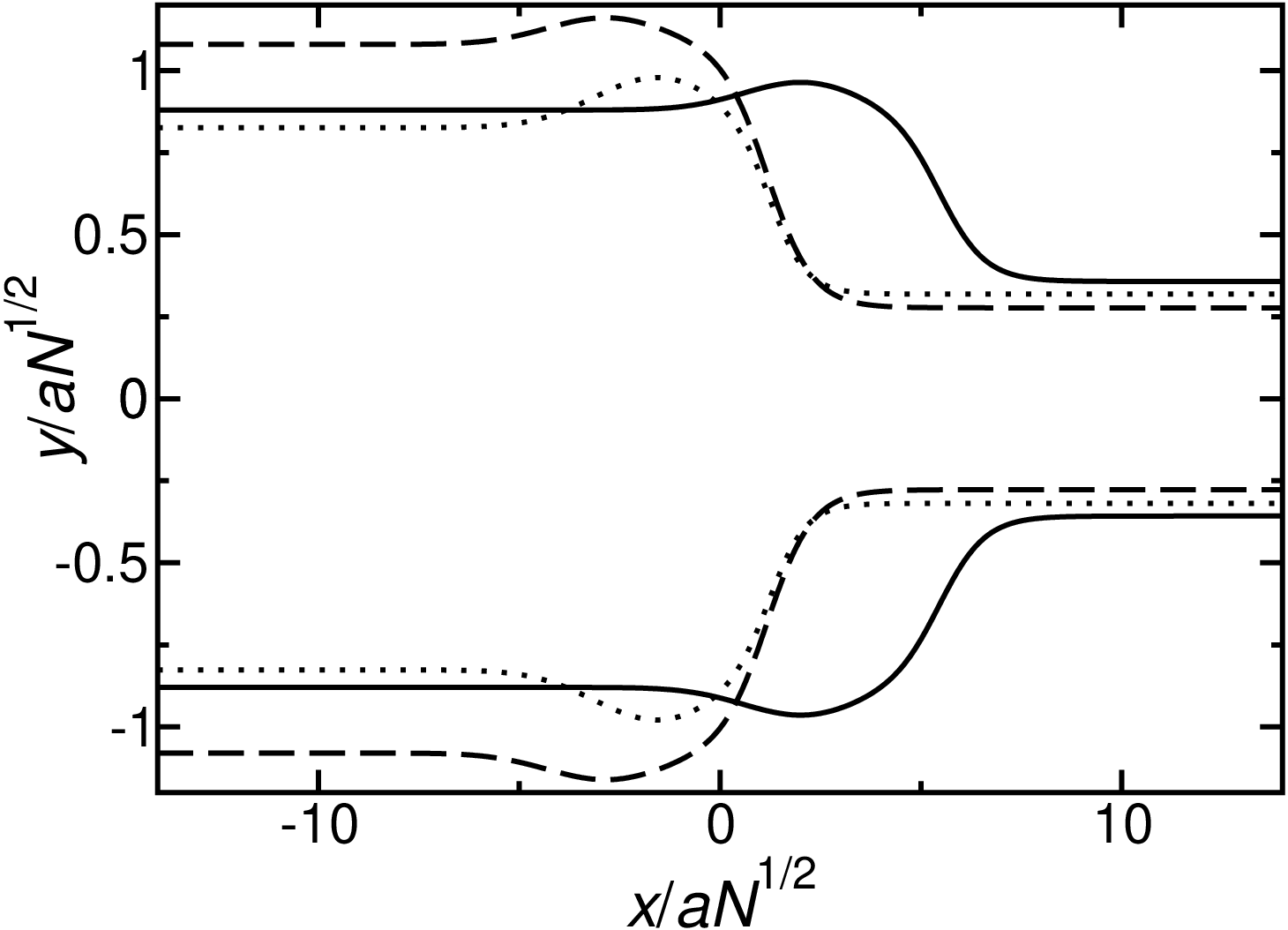}
\caption{\label{outlines_fig} Outlines of the hydrophobic bilayer core
  for $\omega=0.125$ (dotted lines),
  $\omega=1$ (full lines) and $\omega=4$ (dashed lines). Note the
  difference in scale between the $x$ and $y$ axes.} 
\end{figure}

These
results are shown in Figure \ref{outlines_fig}. The dotted line shows the
outline of the membrane core when the oil molecules are very short,
with $\omega=0.125$. Here, the core profile has a noticeable lip
region on the left-hand side of the bilayer just before the domain
boundary, which is located close to the centre of the system. As the size of the oil molecules is increased, so that
$\omega=1$, they are pushed into the thicker side of the bilayer, as already
seen in Figure \ref{profiles_fig}b. We then obtain the core profile plotted
with a full line in Figure \ref{outlines_fig}, where the area of the 
left-hand domain has increased and the domain boundary has moved
to the right.

The plots of the core outlines now bring out some
features of this phenomenon that were not apparent from the cuts
through the density profiles shown in Figure \ref{profiles_fig}. First, we
see that the thickness of the left domain increases relatively
little as $\omega$ is increased from $0.125$ to $1$. This is because,
although the oil molecules are now longer than the hydrophobic
components of the shorter amphiphiles and can no longer easily be
accommodated on the right-hand side of the bilayer, they are still
sufficiently short to mix well
with the corresponding sections of the larger amphiphiles without
causing the left-hand side of the bilayer to thicken. The left domain
then accommodates the extra oil by increasing its area rather than its
thickness, leading to the shift of the domain boundary to the right
noted earlier. In addition, we see that this has the effect of smoothing the surface of the
bilayer core, reducing the size of the lip feature just before the domain
boundary, and also reducing the slope of the core profile $y(x)$ at
the boundary itself.

As $\omega$ is increased still further, to a
value of $4$, we obtain the core profile plotted with a dashed line in
Figure \ref{outlines_fig}. In this case, the oil molecules are almost all
located on the left. Furthermore, their length means that they now
increase the thickness of the left-hand domain. This means that it is now no longer
necessary for this domain to grow in area in order to accommodate the
oil molecules, and the interface is again found close the centre
of the system.

\begin{figure}[h]
\includegraphics[width=\linewidth]{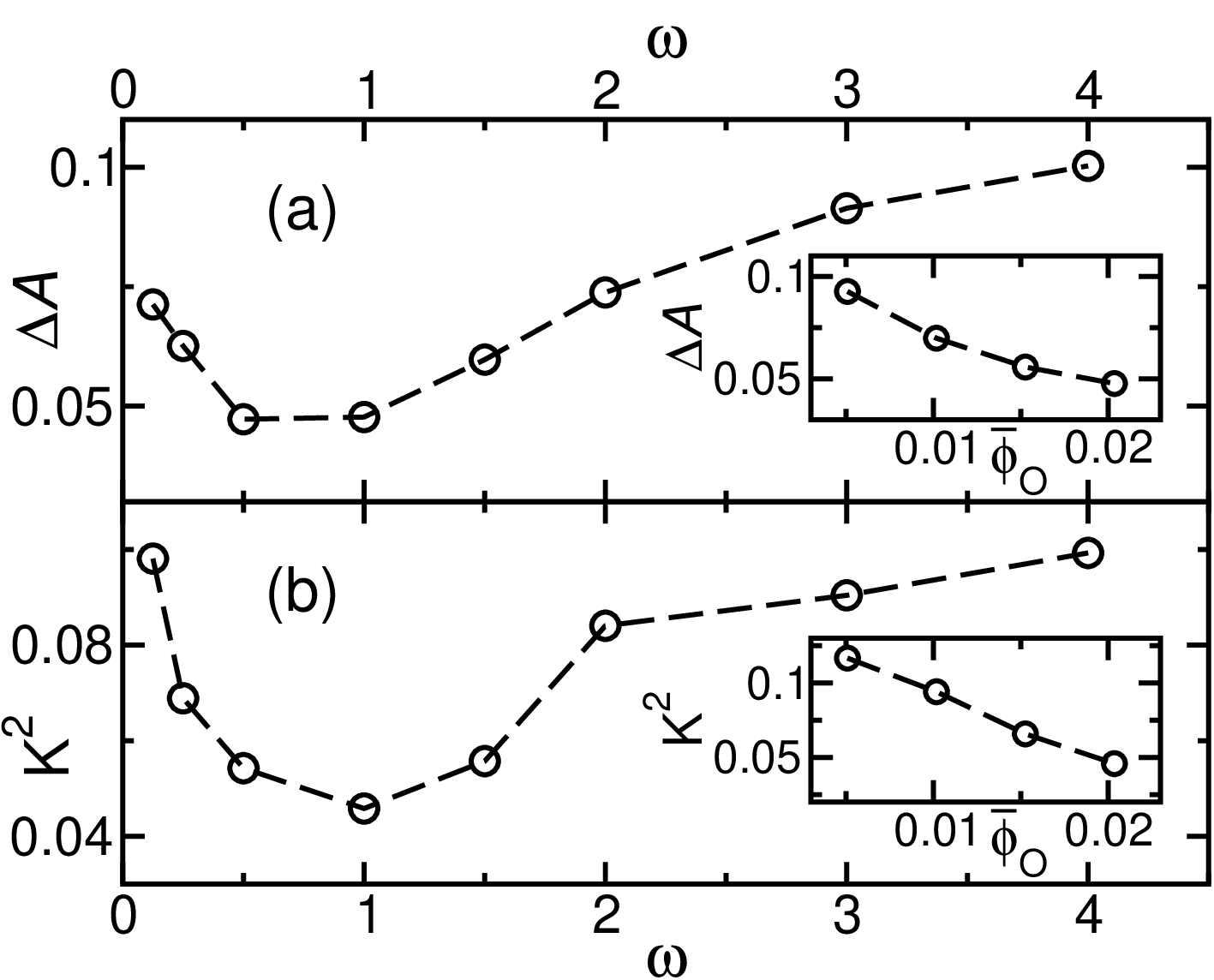}
\caption{\label{excess_curvature_fig} (a) Excess surface area of the
  hydrophobic bilayer core plotted
  as a function of oil size. Inset shows excess surface area plotted
  against oil concentration for $\omega=1$. (b) Corresponding plots of
  the integrated squared
  curvature of the hydrophobic bilayer core.}
\end{figure}

In our discussion of the bilayer core outlines plotted in Figure \ref{outlines_fig},
we noted several effects of varying the size of the oil molecules:
changes in the bilayer thicknesses, the domain sizes and the
structure of the interfacial region. We now wish to proceed to a more
quantitative analysis of the core shape. First, by comparing the core
outlines for $\omega=0.125$ and $\omega=4$ (dotted and dashed lines
respectively), we see that the significant difference in the left-hand
domain thickness between
the two bilayers leads to a larger surface area of the
bilayer core when $\omega=4$, visible as an increase in the length of the outline
plotted in Figure \ref{profiles_fig}. This increases the contact
between the hydrophobic core and the solvent, leading to a sharp
increase in the free energy \cite{seifert_rev}. To quantify the
differences in core surface area between the different bilayers, we
calculate the excess area
\begin{equation}
\Delta A=\int\mathrm{d}x\left[\sqrt{1+(\mathrm{d}y(x)/\mathrm{d}x)^2}-1\right]
\label{excess_area}
\end{equation}
in each case, using the fits to our SCFT data given by Equation
\ref{boltzmann_gaussian}, and plot the results as a function of $\omega$ in Figure
\ref{excess_curvature_fig}a. Since the two bilayer domains are flat,
the major contributions to $\Delta A$ come from the boundary
region. Calculating $\Delta A$ will therefore give us insight into the
free-energy penalty incurred by the introduction of a domain boundary
into the system.

From Figure \ref{excess_curvature_fig}a, we see at once that the
excess surface area $\Delta A$ has a clear minimum at $\omega\sim
1$. This is a result of two of the effects discussed above. The
transfer of oil to the left domain as $\omega$ is
increased initially leads to a lateral expansion of this region and a
smoothing of the lip feature, resulting in a fall in $\Delta
A$. However, as $\omega$ is increased further, the difference in
thickness between the two domains grows, causing an
increase in $\Delta A$. These results show that there is an optimum
oil size at which the free energy penalty arising from the excess
surface area can be minimised in our system.

The lowest free-energy state of a symmetric bilayer is flat,
and deviations from this shape will lead to an increase in its free
energy \cite{safran_pra}. These deviations can be characterised by the
curvatures of the membrane leaflets. Although our laterally segregated
membrane is more complex than a bilayer vesicle or a monolayer
of surfactants at an oil-water interface, situations which can be
studied in detail by models based wholly on membrane curvature
\cite{safran_pra}, study of its surface curvature should still give
insight into how the addition of oil molecules of various sizes pushes
the bilayer into more or less favourable configurations. Since the
core outlines have the form $y=y(x)$, we can calculate the squared
curvature integrated over the bilayer from $x=-L_x$ to $x=+L_x$ using
\begin{equation}
K^2=\int^{+L_x}_{-L_x}\,\mathrm{d}x\frac{(\mathrm{d}^2y(x)/\mathrm{d}x^2)^2}{[1+(\mathrm{d}y(x)/\mathrm{d}x)^2]^3}
\label{kappa_squared}
\end{equation}
In Figure
\ref{excess_curvature_fig}b, we plot $K^2$ as a function of oil
size. As in the case of the excess surface area, we see a clear
minimum around $\omega\sim 1$, where the oil molecules move into the
thicker bilayer domain and smooth the surface. The form of the curve
is slightly different than that seen for $\Delta A$, with $K^2$
increasing rapidly between $\omega=1$ and $\omega=2$ before levelling
off somewhat for $\omega>2$. The reason for this is that, although the growth
in the difference in thickness between the two domains for $\omega>2$
requires an increase in the surface area of the bilayer core in the
domain boundary region, it does not need a similar increase in its
curvature, since the new surface area in the step region is close to
being flat. This can be seen by looking at the dashed outline
($\omega=4$) in Figure \ref{outlines_fig} and comparing it with the
other two outlines.

\begin{figure}[h]
\includegraphics[width=\linewidth]{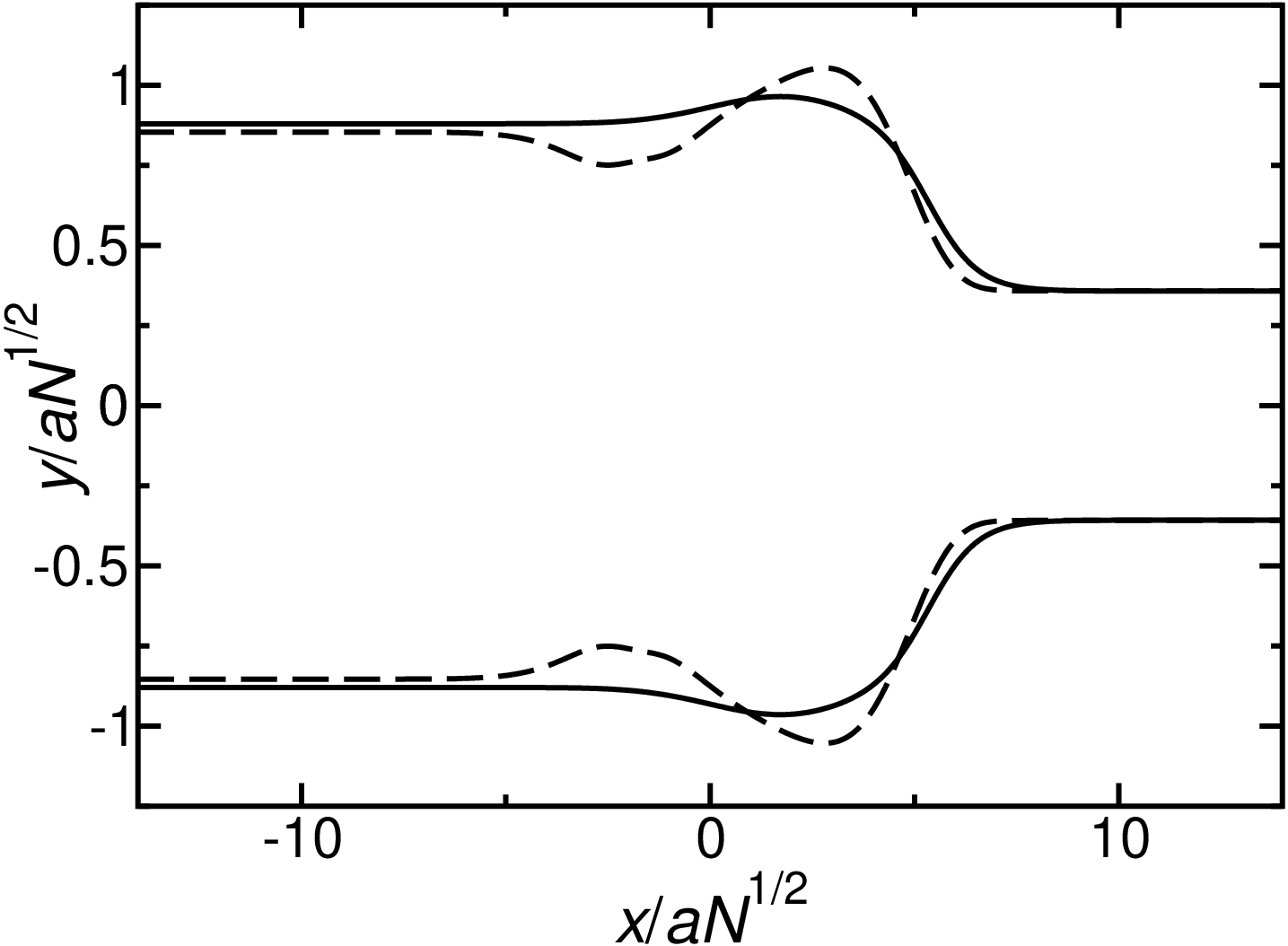}

\caption{\label{vary_oil_fig} Outlines of the hydrophobic bilayer core for
$\overline{\phi}_\text{O}=0.00509$ (dashed lines), and
$\overline{\phi}_\text{O}=0.02036$ (full lines).}
\end{figure}

Finally, we examine the effect of lowering the oil concentration
$\overline{\phi}_\text{O}$ on
the bilayer shape. In the following calculations, when we reduce $\overline{\phi}_\text{O}$ by a
given amount, we increase each of the concentrations
$\overline{\phi}_\text{AB}$ and $\overline{\phi}_\text{AB2}$ by the
same amount. This keeps the total amount of hydrophobic B material
in the system constant, since both the amphiphile species contain
equal amounts of hydrophilic and hydrophobic monomers. The insets to
Figure \ref{outlines_fig}a and b show the effect of this on $\Delta
A$ and $K^2$ respectively for the case of $\omega=1$. Both
quantities rise appreciably as the oil concentration falls.
$\Delta A$ increases by a factor of two as
$\overline{\phi}_\text{O}$ is decreased by a factor of four, and
$K^2$ rises still more, reaching higher values than were obtained
for the original oil concentration even for $\omega=4$.

These increases reflect a change in the bilayer shape: the lip feature near
the domain boundary has become more pronounced. In fact, to calculate
$\Delta A$ and $K^2$ for the lowest oil concentration shown in
the insets to Figure \ref{outlines_fig}, we need to add an extra term
to Equation \ref{boltzmann_gaussian} in order to account for the
increased inhomogeneity in the surface of the bilayer core. We find
a fit of the same high quality as before is obtained if this term
takes the form $a_7\exp[-(x-a_8)^4/a_9]$, where the $a_i$ are
adjustable coefficients. The outline of the bilayer core for the lowest oil
concentration considered, $\overline{\phi}_\text{O}=0.00509$, one
quarter of the original value, is plotted with a dashed line in Figure
\ref{vary_oil_fig}. The corresponding outline at the original
$\overline{\phi}_\text{O}$ is shown for reference, and the difference
in the surface structure between the two bilayers is clear.

The inhomogeneity of the bilayer in the junction region observed in
Figure \ref{vary_oil_fig} reflects an increasing instability in the
bilayer structure. If we set the oil concentration to zero and
increase the amphiphile concentrations so that the total amount of
hydrophobic material remains constant, as described before, the
segregated bilayer structure is no longer stable, and splits into
separate thick and thin bilayers. This shows that, in our system, the
oil is a necessary stabilising factor to overcome the strong size
mismatch between the two amphiphile species.

\section{Conclusions}\label{conclusions}
Using a coarse-grained mean-field model, we have investigated the
effect of added oil on the structure of the boundary between two domains in
a segregated bilayer formed of a mixture of long and short
amphiphiles. We have found that adding oil molecules is a promising
method for controlling the inhomogeneity of the bilayer core surface
in the vicinity of the domain boundary. In particular, we have shown
that, in our model, the surface area of the hydrophobic core exposed to the solvent and
the curvature of the hydrophobic-hydrophilic interface depend in a
very similar way on the length of the oil molecules, and could therefore
be adjusted simultaneously to tune the free energy associated with
boundary formation and hence the size distribution of the
domains. Furthermore, bilayers with added oil are found to be stable
even when a large difference in size between the two amphiphiles leads
the corresponding oil-free bilayer to split.

We now discuss possible extensions to our work that could reinforce
and add detail to
our conclusions. First, we reiterate that the curvature and, particularly, the surface area of the hydrophobic
core in the domain boundary region will be two important factors in
determining the free energy of the boundary, or line tension. In fact,
the
energy cost in changing the surface area of a membrane is so high that bilayer
vesicles have an almost constant area at constant temperature
\cite{seifert_rev}. Any procedure that alters the membrane surface area at the
boundary between two domains therefore has clear potential for controlling the
line tension. However, it would be helpful to back this up by carrying
out a more detailed study of the thermodynamics of the membrane. A
larger system could be used, so that the boundary separates two large
flat domains, allowing the line tension to be separated out directly
from the SCFT free energy by subtracting the energy of the two domains
with a step boundary. The lateral tension in the bilayer
could also be investigated in more detail, as it too will be affected by the changes in surface
area introduced by the oil. For example, a decrease in
surface area could lead to an increase in the lateral tension. The
thermodynamic ensemble could also be chosen
to perform the calculations at constant lateral tension rather than
constant area.

At present, the size mismatch between the two amphiphiles in our
system is very large. This was a deliberate choice in order to bring
out the effects of oil on the bilayer shape as clearly as
possible. However, the size difference in a real system will be
smaller, and it would be very useful to extend our calculations to a
more realistic size ratio. Our assumption of zero repulsive
interactions between the two amphiphile species could also be relaxed,
to allow for a degree of chemical incompatibility. These calculations could involve more detailed modelling
of the amphiphiles, perhaps involving extentions of SCFT beyond the
Gaussian chain approximation \cite{matsen_short}. 

\section{Acknowledgements}
M.J.G. gratefully acknowledges funding from the EU under an FP7 Marie
Curie fellowship and from the IRTG Soft Matter Science.

\end{document}